# Rapid falling of an orbiting moon to its parent planet due to tidal-seismic resonance

Yuan Tian[1], Yingcai Zheng[1]
[1]Department of Earth and Atmospheric Sciences, University of Houston, Houston, TX

**Abstract**

Tidal force plays an important role in the evolution of the planet-moon system. The tidal force of a moon can excite seismic waves in the planet it is orbiting. A tidal-seismic resonance is expected when a tidal force frequency matches a free-oscillation frequency of the planet. Here we show that when the moon is close to the planet, the tidal-seismic resonance can cause large-amplitude seismic waves, which can change the shape of the planet and in turn exert a negative torque on the moon to cause it to fall rapidly toward the planet. We postulate that the tidal-seismic resonance may be an important mechanism which can accelerate planet accretion process. On the other hand, tidal-seismic resonance effect can also be used to interrogate planet interior by long term tracking of the orbital change of the moon.

## 1. Introduction

Darwin (1898) first proposed the idea of a gravity-seismic coupled resonance of a fluid planet to explain the Moon formation and he argued that the violent vibration of the planet at resonance "… *shook the planet to pieces, detaching huge fragments which ultimately were consolidated into the Moon?*". While Darwin's idea of moon formation theory had been largely discarded, very little work has been done to investigate possible consequences of the tidal-seismic resonance for a planet-moon system, in particular for a solid planet.

Tidal force frequencies are usually out of the range of planet free oscillation (or normal mode) frequencies. Even the tidal force of the fastest rotating system in solar system (i.e., Mars-Phobos system) does not have significant effect on the martian free oscillation in the current orbital configuration (Lognonné et al., 2000).

However, in some cases can tidal force frequencies intrude into the frequency range of the planet normal modes. For example, the tidal force on a rapid rotating planet can excite normal modes of the planet (Braviner and Ogilvie, 2014a, b; Barker et al., 2016). Interaction between Saturn ring and Saturn can excite acoustic free oscillation of Saturn (Marley, 1991; Marley and Porco, 1993; Marley, 2014). Fuller (2014) used the tidal force to detect the acoustic free oscillation frequencies from observing density wave in the Saturn ring.

Our goal here is to do a theoretical and numerical analysis to investigate some first order effects of the tidal-seismic resonance.

## 2. Material and Methods

### 2.1. Model setup

In our analysis, we make some simplifications and approximations. We consider a planet-moon system as a binary rotating system in an inertial reference frame. We assume the moon as a point mass and we do not consider potential fragmentation of the moon at the Roche limit (Aggarwal and Oberbeck, 1974; Asphaug and Benz, 1994; Black and Mittal, 2015). The planet does not spin with respect to the reference frame. We assume that the moon's orbit is circular on the planet's equatorial plane. The orbiting period of the moon can be computed using the Kepler's law. We compute the moon's tidal force for every point in the planet by subtracting the centrifugal force from the gravitational attraction force.

We consider two planet models, model-1 which has no topography and model-2 with topography. In model-1, the planet is a homogeneous elastic spherical solid with no topography. We set the compressional wave velocity in the solid as $v_p = 3$ km/s and shear wave velocity $v_s = 1.2$ km/s. We set the radius of the planet as 2000 km. We use the mass-radius relation (Chen and Kipping, 2017) to set the planet density as $\rho = 2.84 kg/m^3$. The mass of the moon is taken as, $10^{16}$ kg (as a reference, this is similar to the mass of an object like Phobos), which is about $10^{-7}$ times of the mass the planet. We also consider the effect of $Q$, which captures the dissipation effect of the planet. Previous work showed that $Q$ could cause tidal phase lag and was important in calculating the orbital decay of moon (Zharkov and Gudkova, 1997; e.g., Bills et al., 2005; Nimmo and Faul, 2013; Zheng et al., 2015). The planet topography can also play a role in the tidal-seismic resonance. Therefore, we also consider a second planet of the same material but with a randomly generated topography (named as model-2).

## 2.2. Method

To see when the resonance can occur, we can compute the planet normal-mode frequencies (see Appendix-A) and the tidal forces (see Appendix-B). Because the moon orbits around the planet periodically, the tidal force is periodic for any location in the planet. If the orbit frequency is designated as $\omega_0$, it can also excite higher order harmonics at integer frequencies such as $n\omega_0$ where n=2, 3, 4, … The tidal-seismic resonance occurs when a tidal force frequency is the same as a normal-mode frequency (Figure 1). Tidal force preferentially excites the fundamental spheroidal normal mode, $_0S_n$, where *n* is the degree in the surface spherical harmonic function.

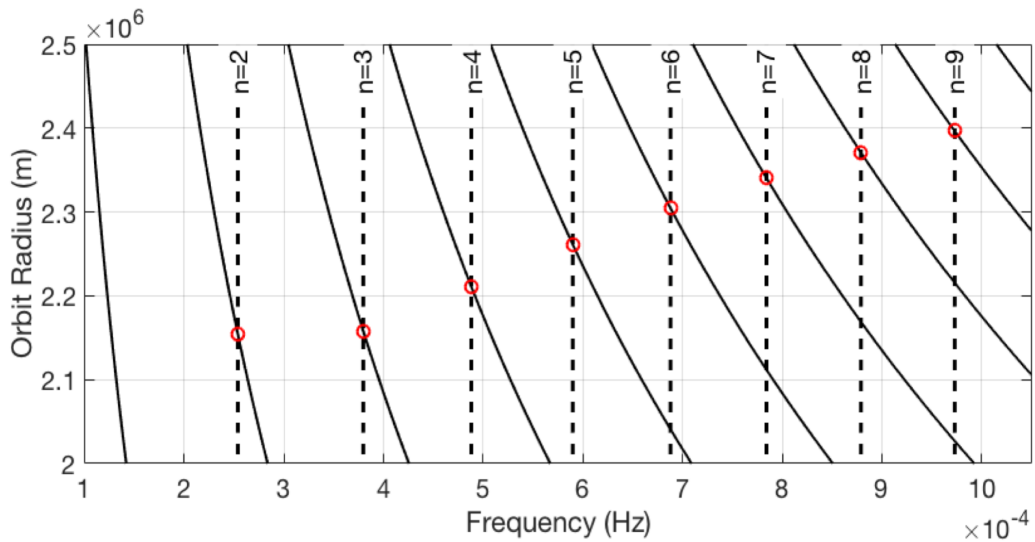

Figure 1. Tidal force and planet normal mode frequencies for model-1. Solid lines are tidal force frequencies calculated at a location on the planet equator. Vertical dash lines are planet free-oscillation frequencies for the spheroidal $_0S_n$ normal modes (n=2, 3, 4 …). Tidal-seismic resonance occurs when tidal force frequencies intersect the planet free oscillation frequencies at red circles.

The orbiting moon exerts a cyclic tidal force for every point in the planet. Therefore, this tidal force can cause seismic displacement in the planet to deform the planet figure. The change of figure of the planet can change the planet gravitational field to exert a net torque on the moon. To compute this torque, we need to compute the tidal-force induced seismic wavefield and we use the boundary element integral equation approach (Zheng et al., 2016a) to do so (see Appendix C). The advantage of this computational method is that it is implemented in the frequency domain and can model long-term seismic field evolution (i.e., can avoid numerical dispersion in many time-domain methods) and can also handle planet topography. We did not include the effect of self-gravitation on the seismic waves (Dahlen and Tromp, 1998) in our calculation. Once we obtain the seismic wavfield we can compute the time dependent torque on the moon and the orbital decay rate of the moon (See Appendix D).

## 3. Results

### 3.1. Tidal-seismic resonance at low orbits

At low orbits, it is evident that the orbital decay rate overall is increasing as the moon is approaching the planet and the tidal seismic effect punctuates/accelerates this trend locally at several distinct orbital radii ("peaks" in Figure 2). These are special orbital radii at which strong torque has been exerted on the orbiting moon due to the resonance. Because of their specialty, we call these orbital radii, $r_n^*$, where $n = 1, 2, 3 \ldots$. At $r_n^*$, the planet $_0S_n$ normal mode frequency is exactly $n$ times of the orbital frequency at $r_n^*$. In these cases, both the tidal force and the $_0S_n$ mode have degree-$n$ spatial patterns in the planet. Therefore, this is a simultaneous coupling between the temporary and spatial frequencies (i.e., degree-$n$) at $r_n^*$, which can cause very strong excitation of seismic displacements in the planet (Figure 2).

We compute orbital decay rates for several different Q values. Large *Q* values correspond to large orbital decay rates because the induced seismic surface displacement and torque are large (Figure 2). In particular, at $r_2^*$ where the excited normal mode is the gravest "football-shaped" mode, the orbital decay rate is on the order of ~1-10 cm/s for different *Q* values (Figure 2) for this particular model. The important thing to note is the relative orbit decay rate for this model. At the tidal-seismic resonance orbit, the orbit decay rate is 2 orders of magnitude more than that for a neighboring orbit where there is no resonance.

We note that the exact numerical value for orbit decay may vary from model to model. However, the tidal-seismic resonance can significantly accelerate the orbital decay ("peaks" in Figure 2 compared to the smooth background trend). The greater the Q value, the sharper the peaks are. To verify whether these peaks are indeed caused by tidal-seismic resonance, we analytically calculated the normal mode $_0S_n$ frequencies of the planet (See Appendix A). And then, we calculate the corresponding $r_n^*$ whose orbital frequency is 1/*n* times of the $_0S_n$ frequency. We found the calculated $r_n^*$ based on the $_0S_n$ frequency correspond to the peaks of the moon orbital decay rate (Figure 2). In conclusion, the rapid falling of the moon is caused by tidal-excited normal modes of the planet. When the moon's orbit radius is large (e.g., > $2.5 \times 10^6$ m), tidal-seismic resonance effect is not obvious. In this case, smaller Q gives larger orbital decay rate and this is consistent with the tidal drag due to anelasticity effect (Bills et al., 2005).

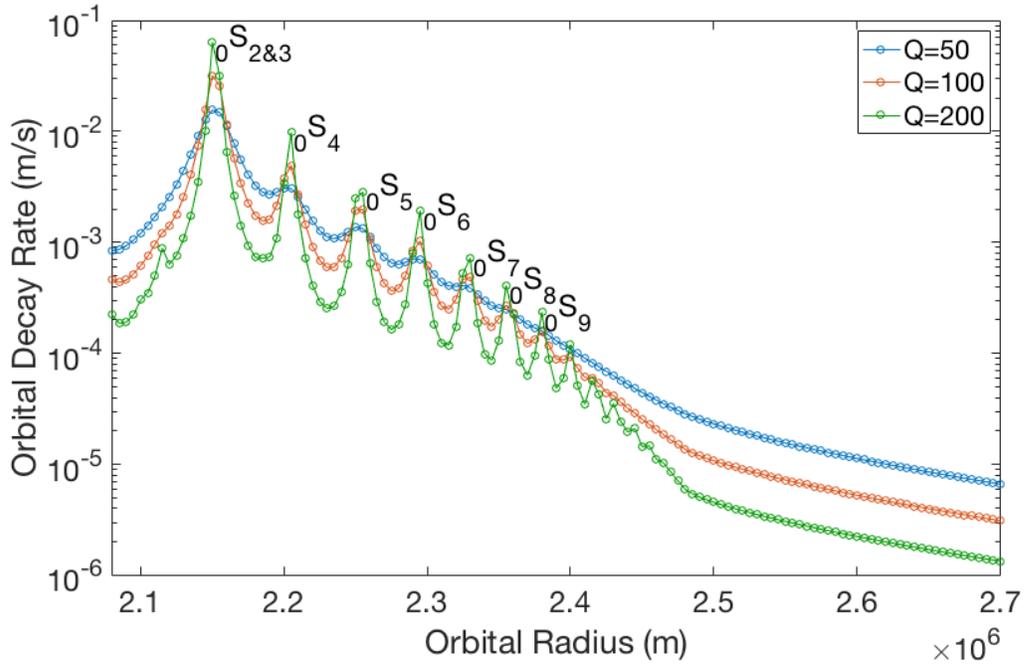

Figure 2. Calculated moon orbital decay rate at different orbital radii and for different Q values for model-1. The horizontal axis is the orbital radius of the moon. The radius of the planet is $2 \times 10^6$m. We also label excited normal modes, $_0S_n, n \geq 2$, expected to be seen for the tidal-seismic resonance.

### 3.2. Topography induced tidal-seismic resonance

At higher orbits, (orbital radius greater than $2.5 \times 10^6 m$ in this case), resonance can occur when $m\omega_0$ matches the $_0S_n$ frequency where both $m$ and $n$ are integers but usually $m > n$. In this case, spatial frequencies of the normal mode and the tidal force de-couple.

In principle, a degree-$m$ tidal force field cannot excite degree-$n$ normal mode for a purely spherical planet (i.e., model-1) because these two fields are orthogonal to each other in space. However, if the planet is not spherical (i.e., model-2), the tidal-seismic resonance can still exist because topography could couple modes of different degrees. Our purpose here is to investigate the topography-induced tidal-seismic effect.

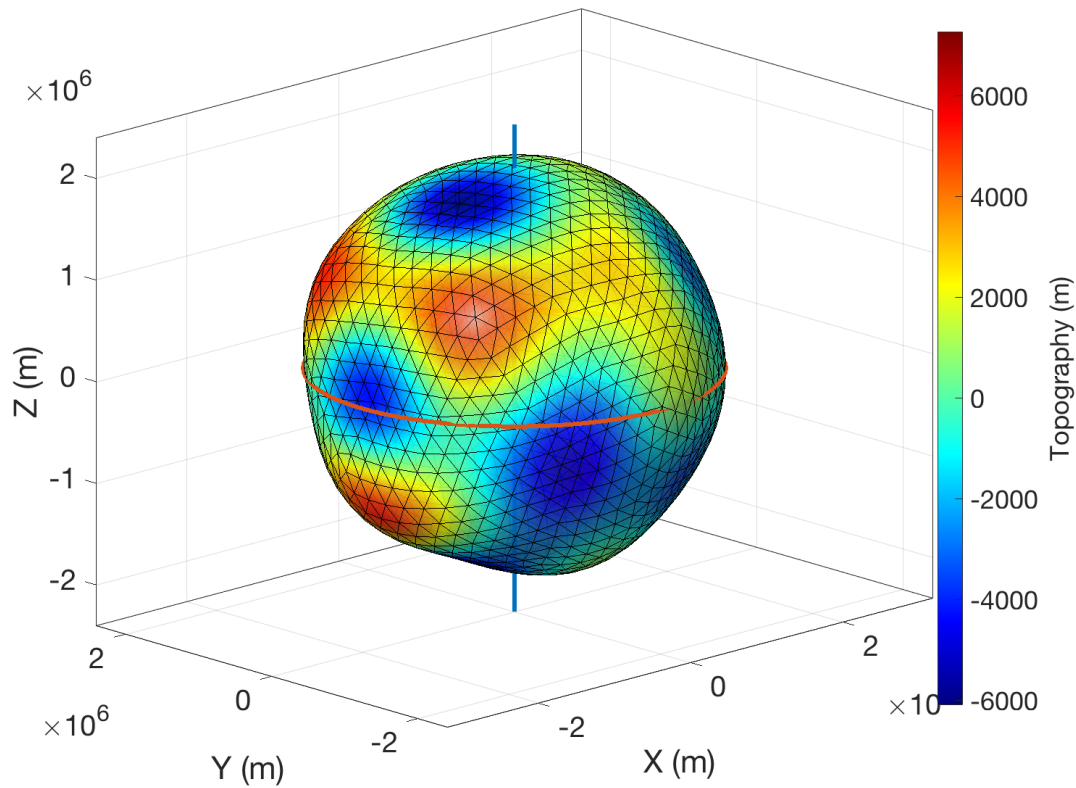

Figure 3. A topography model (model 2) for the planet. We have exaggerated the plotting of the topography by 20 times. Colors indicate topography relative to the mean radius of the planet and the unit of the size is in meters. The thick red line is the equator. The blue vertical thick line is the axis through the poles.

For model-2, we randomly generated a topography for the planet and filtered it spatially using spherical harmonics up to order 6 (including 6) (Figure 3). We use the same numerical procedure laid out in Section 3.1 to compute the seismic wavefield for model-2 and compute the torque on the moon and then the orbital decay rate of the moon. To see the topography effect on the tidal seismic resonance, we can take derivative of the orbital decay rate with respect to the orbital radius and we find localized changes at some radii (Figure 4a). These changes are actually due to topography induced tidal-seismic resonance. To validate this claim, we run our wavefield modeling on two models (model-1 and model-2) and obtain histories of orbital decay rates for both models. We calculate the derivative of the orbital decay rate with respect to orbital radius for the two models (Figure 4 a &b). We see localized changes for the topography model, model-2 (Figure 4a). In contrast, we see a smooth curve of orbital decay rate derivative on model-1 without topography (Figure 4b). Hence, topography can induce tidal-seismic resonance at higher orbits.

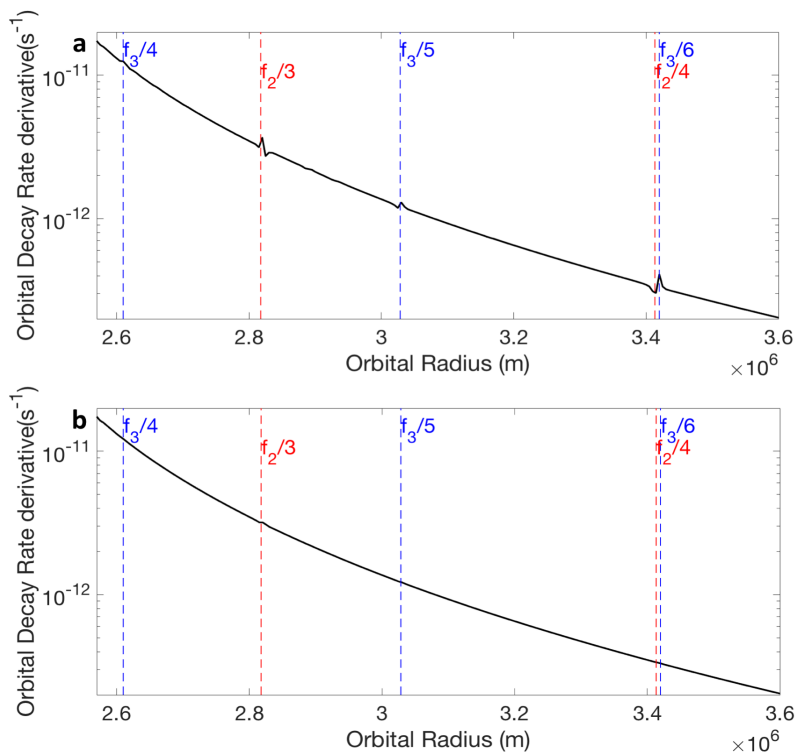

Figure 4. Derivative of the orbital decay rate with respect to orbit radius. (a) for model-2 with topography; (b) for model-1 without topography. Here, $f_n$ is the frequency of the normal mode $_0S_n$. A dashed line shows the orbit whose orbital frequency is $f_n/m$, (n and m are integers) labeled next the dashed line.

## 4. Discussions

As we know, tidal force plays an important role on the orbital evolution of Mars moon, Phobos (Black and Mittal, 2015; Hesselbrock and Minton, 2017). Due to tidal torque, Phobos is spiraling towards Mars. Will the tidal-seismic resonance play a role in this system? At present time, Phobos is too far from inducing a significant tidal-seismic resonance. However, when Phobos' orbit decays to about 1.97 (current orbit is about 2.77) times of the martian radius, topography induced tidal-seismic resonance will be able to cause an orbit decay, estimated about $10^{-10} m/s$ based on current martian topography and an approximate homogenous Mars model (P-wave velocity $v_p = 7.4 km/s$, S-wave velocity $v_s = 3.6 km/s$, density $\rho = 4000 kg/m^3$ to match a $_0S_2$ period of about 2300 s (Zheng et al., 2015)). This rate is then comparable to the current orbital decay rate of Phobos, which is about $1.28 \times 10^{-9} m/s$ (Bills et al., 2005). As Phobos keeps falling towards Mars, the effect of tidal-seismic resonance will be more and more influential to pull Phobos to Mars. Because of the tidal-seismic resonance, we note that Phobos cannot stay at low orbits for long time even it passes the Roche limit and is not fragmented by Mars tidal force.

## 5. Conclusions

Tidal-seismic resonance effect can be important in understanding planet-moon evolution. The tidal seismic resonance can result in a large negative torque, which can increase the orbital decay rate of the moon toward the planet it is orbiting by one order of magnitude. It can also significantly accelerate the accretion process. The tidal-seismic phenomenon may also provide us a potential possible way to interrogate structure and composition information of a planet without having to land an instrument on the surface of planet. By precisely measuring the orbital decay rate as a function of radii, we can then infer the normal modes frequencies of the planet, which can tell a wealth of information about the planet interior.

**Declarations of interest:** none

# Appendices

## A.    Normal-mode frequencies of a homogenous solid planet

A planet can resonate as a whole at certain discrete frequencies and spatial patterns. These vibrational modes are called seismic normal modes (p. 337, Aki and Richards, 2002). There are two types of modes: spheroidal modes associated with volumetric changes and toroidal modes not associated with volumetric changes. Thus, only spheroidal modes contribute to the tidal torque for a homogenous spherical solid planet.

By applying a free surface condition, we can analytically solve for frequencies of the spheroidal modes (see p. 364, Ben-Menahem and Singh, 1981).

$$\frac{2(1-l+k_\alpha a S_\alpha)}{[-2+2l^2-(k_\beta a)^2+2k_\beta a S_\beta]} - \frac{[2l(l-1)-(k_\alpha a)^2 \tau^2 + 4k_\alpha a S_\alpha]}{2l(l+1)[(l-1)-k_\beta a S_\beta]} = 0 \ , \qquad (A.1)$$

where $a$ is the planet radius, $j_l$ is the spherical Bessel function of the angular order $l$ ($l$=0,1,2…), $S_\alpha = j_{l+1}(k_\alpha a)/j_l(k_\alpha a)$, $S_\beta = j_{l+1}(k_\beta a)/j_l(k_\beta a)$, $k_\alpha$ and $k_\beta$ are the P and S wavenumbers respectively, and $\tau = k_\beta/k_\alpha$. The temporary frequencies are included in wavenumbers as, $k_\alpha = \frac{\omega}{v_p}, k_\beta = \frac{\omega}{v_s}$. As an example, we can solve equation (A.1) to get the frequencies of spheroidal modes (Figure A.1) with parameters, $a = 2000 km$, $v_p = 3 km/s$, $v_s = 1.2 km/s$.

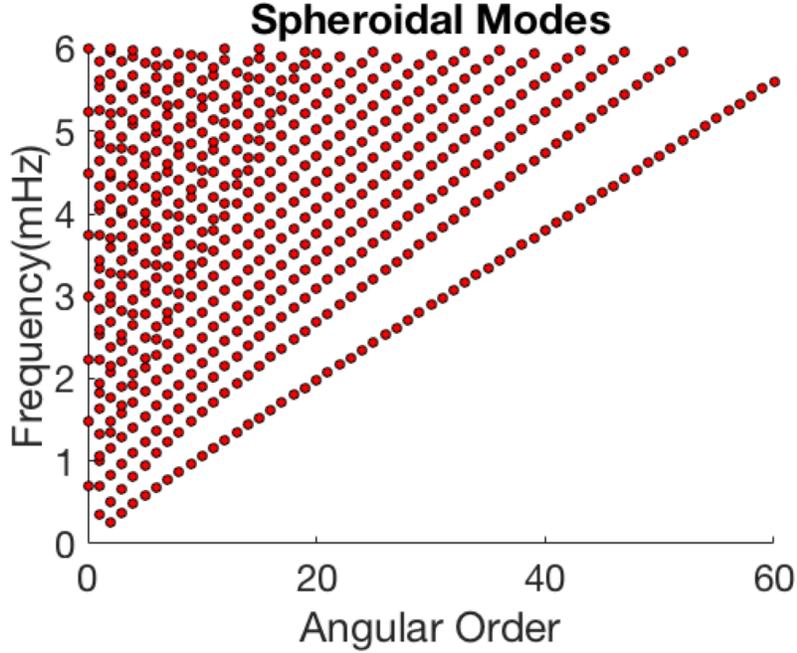

Figure A1: The normal-mode frequencies of the planet model-1. This plot shows the frequencies of spheroidal modes at different angular order.

## B.  Tidal force of a moon

The tidal force of a moon is following Newton's law of universal gravitation. Thus, we can obtain the formula of the tidal force acceleration, **g**, at $\mathbf{x}'$:

$$\mathbf{g}(\mathbf{x}',t) = \frac{Gm_{moon}}{\left|\mathbf{x}_m(t)-\mathbf{x}'\right|^3}[\mathbf{x}_m(t)-\mathbf{x}'] + \mathbf{f}_{cf}(\mathbf{x}',t) , \qquad (B.1)$$

where $\mathbf{x}_m(t)$ is the position of the moon at time $t$ on a circular orbit. The angular speed, $\omega_0$, is assumed to be constant. $G$ is the universal gravitational constant. $m_{moon}$ is the mass of the moon. $\mathbf{f}_{cf}(\mathbf{x}',t)$ is the centrifugal force at $\mathbf{x}'$. We can use this formula to calculate the incident field for our wavefield simulation.

Although the moon's orbital frequency is $\omega_0$, it can also generate other higher harmonic frequencies. To see this, we can analyze the first term in (B.1) by looking at its gravitational potential (Taylor and Margot, 2010)

$$V(\mathbf{x},t) = -\frac{Gm_m}{|\mathbf{x}_m(t)-\mathbf{x}|} = -\sum_{n=1}^{\infty} \frac{Gm_m |\mathbf{x}|^n}{r_m^{n+1}} P_n[\cos\psi(t)], \tag{B.2}$$

where $P_n$ is a n-th order Legendre polynomial, $\psi(t)$ the angle between vectors $\mathbf{x}$ and $\mathbf{x}_m(t)$, and $r_m$ the radius of the moon's circular orbit in the planet equatorial plane. For a point $\mathbf{x}$ in the planet, the gravity potential is changing with time due to the term in the Legendre polynomial $P_n[\cos\psi(t)]$. If the orbit frequency is $\omega_0$, $P_n[\cos\psi(t)]$ has the term $\cos(n\omega_0 t)$, whose frequency is $n\omega_0$ (n=0, 1, 2, ...). So, the tidal force frequencies of a moon are discrete and are higher order harmonics.

In the examples in the paper, we truncate the angular order to $n=16$ in our numerical simulation. We have tested the higher order of tidal force term has no significant influence; thus, we ignored the higher order polynomials. We consider the case that the mass of the moon is much smaller than the planet; therefore, the center of the planet is almost the same as the mass center of the two-body system.

**C.     Boundary element method modeling of seismic wavefield in a planet with topography**

We apply the Boundary Element Method (BEM) (Zheng et al., 2016b) to numerically model the seismic field in the planet excited by tidal force. This method can solve the problem with arbitrary boundary shape in the frequency domain, which reduces the computational cost in this study. The BEM method computes the whole field, including all possible normal mode coupling among spheroid and toroidal modes. The idea of BEM is that it first solves the wavefield on the boundary. Once the boundary field is known, we can compute the wavefield at any point in the planet using the representation theorem(p. 28, Aki and Richards, 2002). The boundary integral equation governs the seismic field, **u**:

$$\frac{1}{2}u_q(\mathbf{x},\omega) = u_{0q}(\mathbf{x},\omega) - \oiint_S u_i(\mathbf{x}')C_{ijkl}(\mathbf{x}')G_{kq,l}(\mathbf{x}',\mathbf{x},\omega)n_j d\mathbf{x}'^2; \mathbf{x},\mathbf{x}' \in S. \tag{C.1}$$

In this equation, S is the surface of the planet including topography; $n_j$ is the outward surface normal along $j$-th direction; $C_{ijkl}(\mathbf{x}')$ is elastic constant matrix of the planet model at $\mathbf{x}'$. Here, we assume the planet is isotropic, so there are only two independent Lame parameters in the matrix. $G_{kq,l}(\mathbf{x}',\mathbf{x},\omega)$ is the elastic green's function derivative with respect to $\mathbf{x}'$ along the $l$-th direction in the frequency domain. All subscripts (q, i, j, k, l) in equation (C.1) take value of 1, 2, or 3 to indicate the component of the vector/tensor field.

The incident field, $u_{0q}(\mathbf{x},\omega)$, excited by the tidal force (see equation (B.1)) along the $q$-th direction in the frequency domain can be computed as:

$$u_{0q}(\mathbf{x},\omega) = \int \rho(\mathbf{x}')g_i(\mathbf{x}',\omega)G_{iq}(\mathbf{x},\mathbf{x}',\omega)d\mathbf{x}'^3, \tag{C.2}$$

where $\mathbf{x}$ and $\mathbf{x'}$ are points in the planet. $G_{iq}(\mathbf{x},\mathbf{x'},\omega)$ is the Green's function in a homogeneous unbounded elastic medium. $g_i(\mathbf{x'},\omega)$ is the tidal force of the moon at $\mathbf{x'}$ along the i-th direction in the frequency domain. The value of $g_i(\mathbf{x'},\omega)$ at each frequency are the Fourier coefficients of equation (B.1):

$$\mathbf{g}(\mathbf{x'},\omega) = \frac{1}{T}\int_0^T \mathbf{g}(\mathbf{x'},t)e^{i\omega t}\, dt\ , \tag{C.3}$$

where $T$ is the orbital period. The frequencies $\omega$ are discrete numbers ($n\omega_0$) as showed in equation (B.2), where $n = 1,2,3\ldots$.

To solve equation (C.1), we partition the planet surface into small triangular elements We assume on each surface element, the seismic wavefield $\mathbf{u}(\mathbf{x},\omega)$ is constant. We then can get a system of coupled linear equation. Equation (C.1) in its discretized form can be written as:

$$\left(\frac{1}{2}I + A\right)[u] = [u_0]\ , \tag{C.4}$$

where $A$ is a matrix representing wavefield interaction between surface elements. Matrix $A$ depends on the geometry of the surface and internal medium properties; $[u]$ is a vector containing 3-component surface displacement for all elements. $I$ is a square identity matrix. $[u_0]$ is a vector containing the incident field on all surface elements excited by the tidal force and is calculated using equation (C.3). In the BEM method, we invert for the surface displacement $[u]$ on each element by solving the linear algebraic equation (C.4). Now, we have obtained seismic displacement $\mathbf{u}(\mathbf{x},\omega)$ on each surface element. By representation theorem (p.28, Aki and Richards, 2002), we can get displacement field at any point within the planet.

Because the tidal force is discrete in frequency; we only need to solve equation (C.4) on those discrete frequencies $n\omega_0$.

Once we have $\mathbf{u}(\mathbf{x},\omega)$ at frequency $n\omega_0$, we can use Fourier series to represent $\mathbf{u}(\mathbf{x},t)$:

$$\mathbf{u}(\mathbf{x},t) = \sum_{n=1}^{16} 2\,\mathrm{Re}[\mathbf{u}(\mathbf{x},n\omega_0)e^{in\omega_0 t}], \qquad (C.5)$$

where Re is to take the real part of a complex number. Because $\mathbf{u}(\mathbf{x},t)$ is linearly proportional to $[u_0]$ and is linearly proportional to the mass of the moon, $\mathbf{u}(\mathbf{x},t)$ is also proportional to the mass of the moon.

### D.    Tidal-seismic torque on the moon and orbital decay rate

Once we obtain $\mathbf{u}(\mathbf{x},t)$, we can compute the time-t dependent torque, on the moon, due to the seismic wavefield:

$$\mathbf{M}(t) = \oiint_S r\mathbf{r}_m(t) \times \mathbf{g}[\mathbf{x},t]\mathbf{u}(\mathbf{x},t)\cdot\mathbf{e}_r\, d\mathbf{x}^2, \qquad (D.1)$$

where, $\mathbf{r}_m(t)$ is the location of the orbiting moon; $\mathbf{g}[\mathbf{x},t]$ is the gravity acceleration exerted on the moon by the excess mass at $\mathbf{x}$; $\mathbf{g}[\mathbf{x},t]$ is proportional to mass of the moon (see Appendix B); $\rho$ is the density of the planet; $e_r$ is the surface normal at $\mathbf{x}$.

We can derive the orbital decay rate of the moon ($\dot{r}_m(t) = dr_m(t)/dt$) by Newton's second law:

$$\left|\dot{r}_m(t)\right| = \frac{2\overline{\mathbf{M}(t)}}{m_{moon}} \sqrt{\frac{r_m(t)}{Gm_{pl}}} \tag{D.2}$$

where $\overline{\mathbf{M}(t)}$ is the average torque on the moon in one orbital period. $G$ is the universal gravitational constant. $m_{moon}$ is the mass of the moon. $m_{pl}$ is the mass of the planet.

We show that both $\mathbf{g}[\mathbf{x},t]$ and $\mathbf{u}(\mathbf{x},t)$ are proportional to the mass of the moon in Appendix B and Appendix C. In equation(D.2), the orbital decay rate is divided by the mass of the moon $m_{moon}$. Thus, the orbital decay rate is proportional to the mass of the moon. If the mass of the moon changes, the orbital decay rate will change accordingly.